\documentclass[twocolumn]{article}

\usepackage{multicol}

\usepackage[T1]{fontenc}
\usepackage{lmodern}
\usepackage{eurosym}
\usepackage{lastpage}
\usepackage{xspace}
\usepackage[margin=15mm,includehead,includefoot]{geometry}
\usepackage{fancyhdr}
\usepackage{booktabs}
\usepackage{graphicx}
\usepackage{multirow}
\usepackage{array}
\usepackage{xcolor}
\usepackage{csquotes}
\usepackage{pgfgantt}
\usepackage{titlesec}
\usepackage{hyperref}
\usepackage{comment}
\usepackage{amsthm}
\usepackage{amsmath}
\usepackage{amssymb}
\usepackage{pifont}
\usepackage{tabto}
\usepackage{enumitem}
\usepackage{multirow}
\usepackage{graphicx}
\usepackage{url}

\usepackage[switch]{lineno}

\titleformat*{\section}{\large\bfseries}
\titleformat*{\subsection}{\normalsize\bfseries}
\titleformat*{\subsubsection}{\normalsize\bfseries}

\renewcommand{\footnotesize}{\fontsize{8bp}{1em}\selectfont}

\DeclareMathOperator*{\argmin}{arg\,min}

\usepackage{authblk}

\begin{document}
\title{\bf Social impact of CAVs -- coexistence of machines and humans in the context of route choice}
\author[1]{Grzegorz Jamr\'oz$^*$}
\author[2]{Ahmet Onur Akman}
\author[2]{Anastasia Psarou}
\author[2 ]{Zolt\'an Gy\"orgi Varga}
\author[1]{Rafa{\l}  Kucharski}
\affil[1]{Faculty of Mathematics and Computer Science, Jagiellonian University, Kraków, Poland}
\affil[2]{Jagiellonian University, Doctoral School of Exact and Natural Sciences}
\setcounter{Maxaffil}{0}
\renewcommand\Affilfont{\itshape\small}
\maketitle

{\bf Abstract}
Suppose in a stable urban traffic system populated only by human driven vehicles (HDVs), a given proportion (e.g. $10\%$) is replaced by a fleet of Connected and Autonomous Vehicles (CAVs), which share information and pursue a collective goal. Suppose these vehicles are centrally coordinated and differ from HDVs only by their collective capacities allowing them to make more efficient routing decisions before the travel on a given day begins.  
Suppose there is a choice between two routes and every day each driver makes a decision which route to take. Human drivers maximize their utility. CAVs might optimize different goals, such as the total travel time of the fleet. We show that in this plausible futuristic setting, the strategy CAVs are allowed to adopt may result in human drivers either benefitting or being systematically disadvantaged and urban networks becoming more or less optimal. Consequently, some regulatory measures might become indispensable.

\begin{figure*}[h!]
\centering
\includegraphics[scale=0.8]{"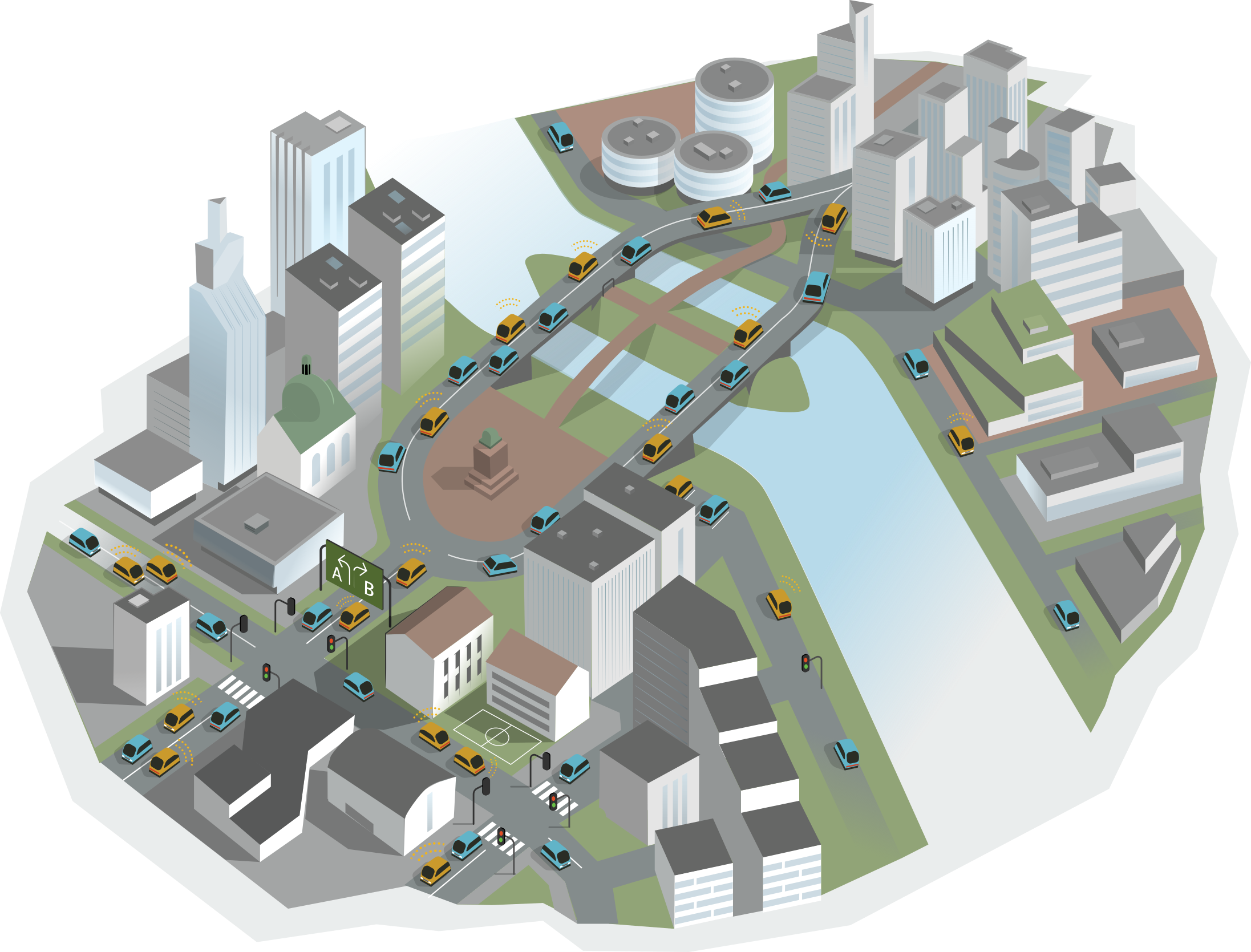"}
\caption{A two-route bottleneck in a city. To reach the other side of the river the drivers have to choose between the alternatives A and B. 
The everyday choice to minimize travel time can be understood as a repeated game between multiple participants striving to find the option which maximizes a driver's utility.}
\label{Fig_1}
\end{figure*}
\begin{figure*}[h!]
\centering
\includegraphics[scale=0.9]{"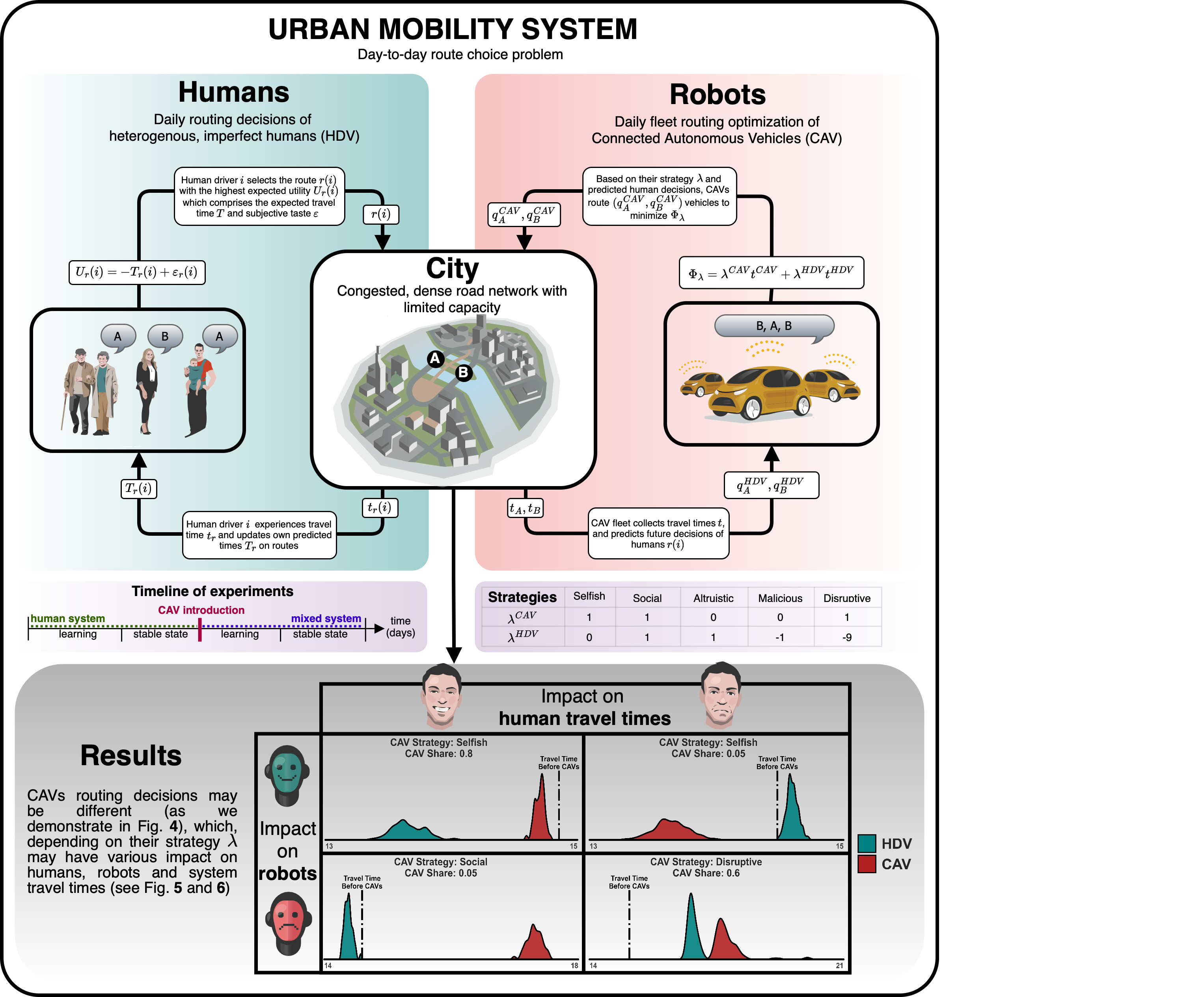"}
\caption{The learning and decision processes applied by human drivers (HDVs) and machines (CAVs). HDVs' reasoning is subjective and based on limited access to information. Contrariwise, CAVs have access to complete information on travel times and make optimal collective routing decisions. The interaction between human agents and CAVs may result in any combination of human drivers and CAVs being better off or worse off subject to the strategy applied by CAVs. In particular, the system-wide welfare may improve or deteriorate in the wake of introduction of CAVs.}
\label{Fig_2}
\end{figure*}

\section{Introduction}
\label{Sec_Intro}
Which route should I take? Millions of people commuting to work by car face this dilemma every day \cite{ABCmobility}. In urban settings the choice is not straightforward as there are usually multiple viable alternatives. In fact, the reasons we select a given route might be very complex \cite{Arthur, Bacharach} ranging from habitual choice or everyday exploration in order to identify the best alternative to anticipating decisions of others. Moreover, people are often very different and might prefer different options in the same situation or behave seemingly irrationally \cite{Glimcher}. Suppose now that in a future urban traffic system with stable drivers' choice strategies a proportion of human drivers (HDVs) is replaced by intelligent vehicles (CAVs) which share information and make collective route choices based on one of the pre-defined collective fleet strategies: 

\begin{itemize}
\item Selfish (minimization of CAVs' \emph{collective} travel time), 
\item Altruistic (minimization of HDVs' mean travel time), 
\item Social (minimization of the mean travel time of all vehicles in the system), 
\item Malicious (aiming to \emph{maximize} HDVs' mean travel time), 
\item Disruptive (maximization of HDVs' travel time at a bounded own cost). 
\end{itemize}
Will, once the system has stabilized again after such disruption, the route preferences of CAVs and HDVs be different?  Will CAVs be better off than the HDVs they replaced? And, crucially, could the human drivers be significantly disadvantaged or the system-wide travel times deteriorate?

In this paper we set out to study these fundamental questions using mathematical models and simulations, see Fig. \ref{Fig_2}. Focusing on the two-route bottleneck settings, Fig. \ref{Fig_1}, which are often present in real systems \cite{Duan, Leclercq}, we discover that:
\begin{itemize}
\item The choices of CAVs that replace a given share of HDVs differ significantly from the choices of the remaining HDVs.
\item In different scenarios the average travel time of both HDVs and CAVs may increase or decrease, Fig. \ref{Fig_2}. 
\item If the fleet of CAVs applies the selfish strategy, it may improve its collective travel time at a cost to human drivers when the share of CAVs is small. 
\item For a large share of CAVs, the selfish or social strategies of CAVs may result in improvement of travel times for all the drivers. This, however, comes at a price of reduced equity.
\item Human driver populations with low perception bias may be less prone to exploitation by intelligent fleets of CAVs than more diverse and less optimal populations. 
\item Heavily congested systems, where the choices of HDVs and CAVs tend to be  similar, may be less susceptible to exploitation by CAVs. Contrariwise, uncongested networks could be easily exploited by machines.  
\item More elaborate, e.g. malicious, CAV strategies may result in oscillations and significant deterioration of driving conditions for all the drivers. 
\end{itemize}
These conclusions seem to have been missing in the literature dealing with CAV - HDV interaction and constitute our original contribution to the subject. We obtain them from simulations by comparing the properties of the two-route bottleneck system before and after the introduction of CAVs.

\begin{figure*}[h]
\centering
\includegraphics[scale=0.45]{"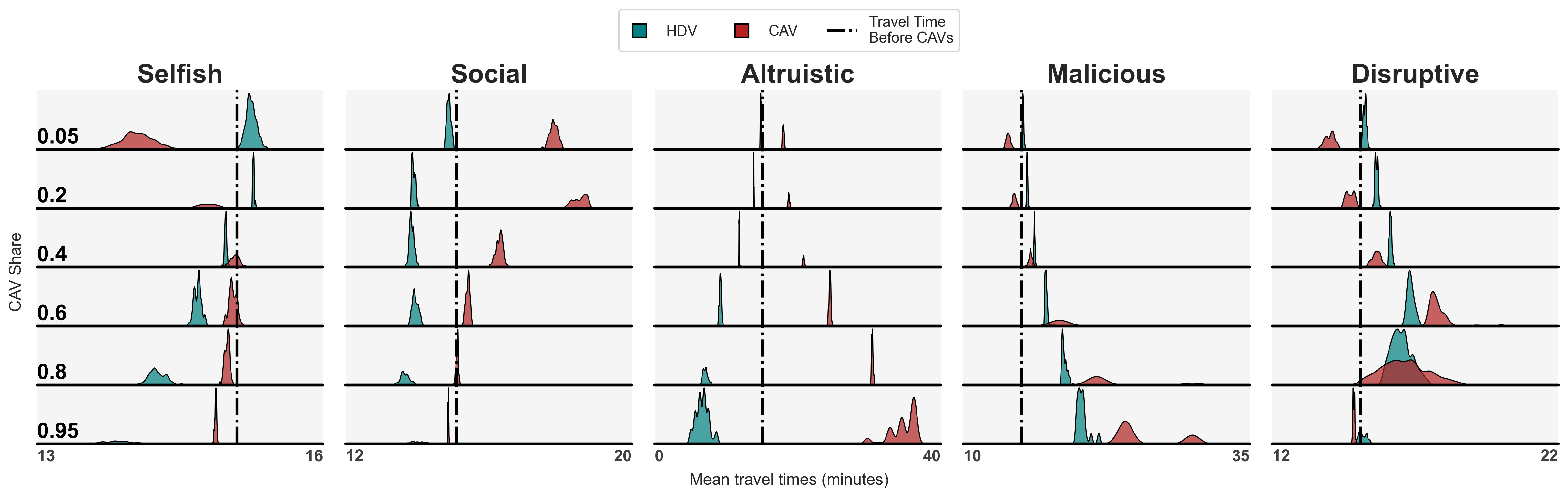"}
\caption{
Kernel density estimations of mean travel times of HDVs and CAVs for different CAV shares and strategies, based on the final $100$ days of the simulation. In the selfish strategy the CAVs experience shorter while HDVs longer travel times for small CAV shares compared to the situation before CAV introduction. For larger CAV shares both groups' travel times improve, with HDVs gaining more. In the altruistic and social strategies the travel times of CAVs increase and those of HDVs decrease, compared to the travel times before the introduction of CAVs except for the social case with very large shares of CAVs when both groups' travel times decrease. The malicious and disruptive strategies are similar to selfish for small CAV shares however they may cause oscillations and lead to increased travel times of all the vehicles for large CAV shares.}
\label{Fig_3}
\end{figure*}

\noindent The standard econometric framework used to quantify choice is the expected utility theory \cite{VonNeumann1944}, which posits that people choose the alternative with the highest expected utility. In the route choice setting with no access to external sources of information, the main component of utility is the predicted travel time \cite{Bogers2007}:
\begin{equation}
\label{Eq_utility}
U_r = -T_r + \mbox {other factors}, 
\end{equation}
where $U_r$ is the utility of route $r$ and $T_r$ is the expected (by a given agent) travel time on route $r$.
If other factors are negligible, the rational HDV choice is  to select the route with the highest utility, which corresponds to the shortest expected travel time. In the case of bottlenecks with two alternatives $A$ and $B$, Fig. \ref{Fig_1}, this amounts to choosing
\begin{equation}
\argmin_{r \in \{A,B\}} T_r.
\label{Eq_Tr}
\end{equation}

\noindent Transport systems analysts typically assume that the system is in or close to equilibrium \cite{Patriksson}. This means that the numbers of drivers traveling along alternative routes within a given time interval, e.g. the morning peak hour,  are stable across consecutive days. This also implies stability of travel times (which may be assumed to depend monotonically, via the BPR \cite{BPR} function, on the number of drivers, see Methods) on different routes. 

\noindent The most classical and widely-accepted traffic equilibrium, postulated by Wardrop \cite{Wardrop}, occurs when no single driver, who is assumed to have infinitesimal influence on the system as a whole, has an incentive to swap routes provided other drivers do not modify their choices the following day. Quantitatively, the drivers are assumed to make choices according to formula \eqref{Eq_Tr}, where $T_r$ are equilibrium travel times
\cite{Wardrop}. This so-called User Equilibrium (UE), is reminiscent of Nash equilibrium \cite{Nash} in game theory and in simple settings can be explicitly computed. When the number of agents is finite, however, the setting becomes an atomic congestion game which is inherently unstable \cite{Ahmad, Harsanyi}, see also Appendix, and often admits multiple Nash equilibria, \cite{Whitehead}. 

A more realistic setting, adopted in our study, assumes that there exist other components of utility in equation \eqref{Eq_utility}, such as tastes or fluctuations in driving conditions, which are incorporated via formula
\begin{equation}
U_r = - T_r + \varepsilon_r,
\label{Eq_Ur}
\end{equation}
where $\varepsilon_r$ are random variables. This setting, the subject of random utility theory \cite{McFadden, Thurstone} implies that, for $\varepsilon_r$ independent identically distributed Gumbel variables (see Methods), the expected proportion of drivers choosing alternative $A$ is given by the logit formula:
\begin{equation}
P_A = \frac {\exp(-T_A\slash \beta)} {\exp(-T_A\slash \beta) + \exp(-T_B\slash \beta)},
\label{eq_PA}
\end{equation}
which is pervasive in transport modelling \cite{CascettaBook}. In \eqref{eq_PA}, $\beta$ is the spread of subjective HDV tastes (perception bias). Low spread corresponds to HDVs preferring routes close to optimal in terms of travel time. High spread makes the choices more random. 
Assuming that the number of vehicles is very large, Daganzo and Sheffi \cite{DaganzoSheffi} postulated the so-called stochastic user  equilibrium (SUE), in which no agent \emph{believes} they can improve their travel time by unilaterally changing routes. 

\begin{figure*}[h]
\centering
\includegraphics[scale=0.45]{"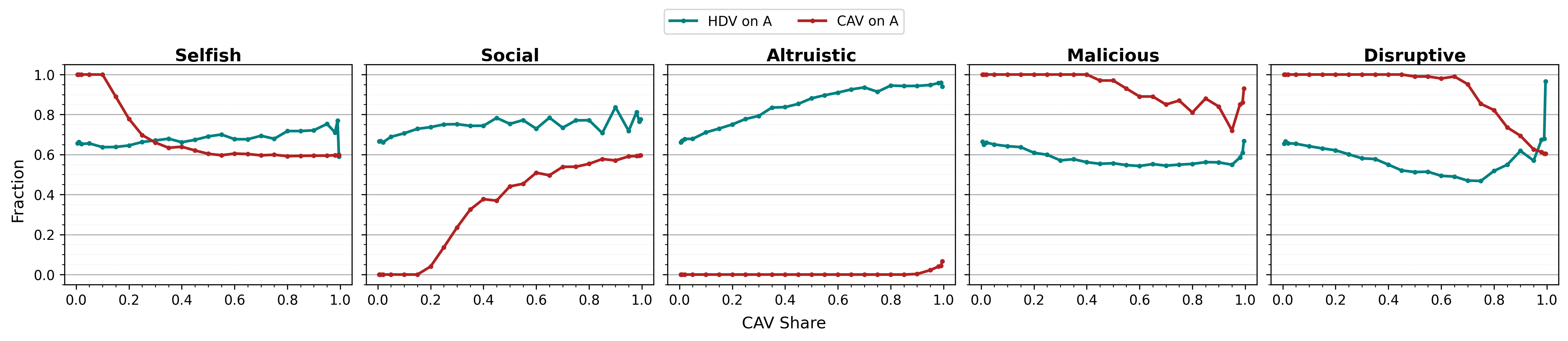"}
\caption{Comparison of average fractions of CAVs and HDVs on route A for different CAV shares. In the selfish scenario, all CAVs are routed via $A$ for small CAV shares and this fraction decreases for increasing CAV share. The fraction of HDVs on $A$ remains relatively stable. In the social scenario the tendency is exactly opposite, i.e. all CAVs are routed via $B$ (corresponding to fraction $0$ on $A$) for small CAV shares. Altruistic CAVs are all routed via $A$ while HDVs strongly prefer $B$. Malicious CAVs behave similarly to selfish CAVs for small shares, however for large shares their strategy entails routing more vehicles, on average, via A. Finally, the disruptive strategy is, in terms of fractions, on average similar to the malicious strategy. }
\label{Fig_4}
\end{figure*}

\noindent Fast-forward to 2024, the logit choice, based on Gumbel-distributed random terms in \eqref{Eq_Ur}, and its variants \cite{ChapterDicreteChoiceModels} is still the most popular family of human route choice models, see also \cite{Daganzo1979, PERTURBED, CULO, WatlingBounded} for other approaches. Accordingly, we adopt a plausible logit-type model, called $\epsilon$-Gumbel, in this paper, see Methods. Importantly, the logit choice formula can be derived not only based on the error in perceived utility as per Daganzo and Sheffi \cite{DaganzoSheffi} but also within the more recent framework of rational inattention \cite{FosgerauRI, MatejkaRI}.
The equilibrium notions of SUE as well as UE, see also BRUE \cite{Di, Mahmassani}, however, seem to be poorly suited to more realistic state-of-the-art models of multi-agent simulations, initiated $40$ years ago by Horowitz \cite{Horowitz} and employed in this paper. Therefore, instead of assuming that the system is strictly in equilibrium, such as SUE, we study experimentally systems which stabilize, see Appendix, bearing in mind that the stable states can be very complex \cite{Cascetta1989, WatlingCantarella} or nonunique \cite{Smith}.

\begin{figure*}[h]
\centering
\includegraphics[scale=0.48]{"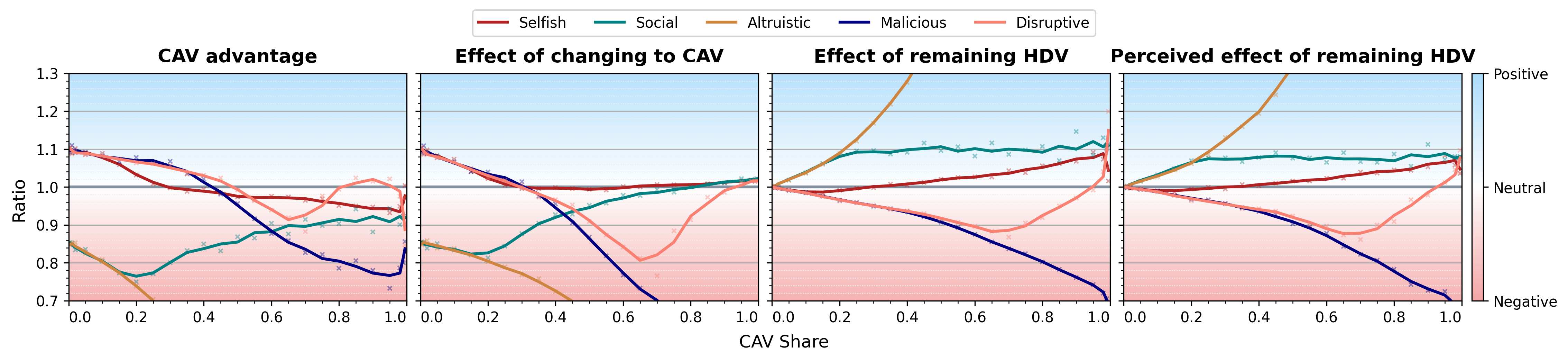"}
\caption{Outcomes of replacing a fraction of HDVs by CAVs for different CAV shares and baseline HDV perception bias. \emph{CAV advantage} $(\tau \slash \rho)$: the ratio of mean HDV travel time averaged over days $301-400$ and mean CAV travel time averaged over days $301-400$. If $\tau \slash \rho>1$ it is better to be CAV than HDV after M-day. \emph {Effect of changing to CAV} $(\tau_b \slash \rho)$ : the ratio of mean HDV travel time averaged over days $101-200$ and mean CAV travel time averaged over days $301-400$. If $\tau_b \slash \rho>1$, the vehicle which switched from HDV to CAV experiences on average shorter travel times. \emph {Effect of remaining HDV} $(\tau_b \slash \tau)$: the ratio of mean HDV travel time averaged over days $101-200$ and mean HDV travel time averaged over days $301-400$. If $\tau_b \slash \tau>1$, the vehicle which remained HDV after M-day experiences on average shorter travel times. \emph {Perceived effect of remaining HDV} $(u_b \slash u)$: the ratio of mean perceived HDV travel time averaged over days $101-200$ and mean perceived HDV travel time averaged over days $301-400$. If $u_b \slash u>1$, the vehicles which remained HDVs after M-day experience on average better perceived travel times.}
\label{Fig_5}
\end{figure*}

As the system is not exactly in equilibrium \cite{CaCa1995}, the drivers do not know precisely the travel times they will experience selecting different alternatives. Therefore, $T_r$ in formula \eqref{Eq_Ur} can only be approximate and we assume that every driver adjusts (in their minds) these estimates every day. There are various mechanism by which the human agents may adjust their day-to-day route choices \cite{Ahmad}. In this paper we only consider the most popular mechanism called, depending on the source, exponential filter or Gawron/Horowitz/Erev-Roth learning \cite{CantarellaBook, ErevRoth, Gawron, Horowitz}, omitting explicit modeling of habitual choice or bounded rationality \cite{CantarellaBook, Mahmassani, WatlingBounded} or direct anticipation of decisions of others based on game theory \cite{Ahmad, Rosenthal}. 
We assume, namely, that every driver maintains implicitly/subconciously estimates of travel times on alternative routes and these estimates are updated daily by combining previous knowledge and most recent travel times. There exist two basic mechanisms, \emph {experience only} and \emph {full information} as well as a whole spectrum of models, where only partial information is available \cite{Friesz, MatejkaRI}. In this paper, we focus on the \emph{experience only} mechanism, in which human drivers' knowledge is updated based on the experienced travel times only and there is no access to past or real-time travel times on alternative routes.

\noindent In our simulations, the human-only system  stabilizes as a result of human learning and adjustment. Once this has happened, a given share of HDVs is replaced by a fleet of CAVs which is centrally controlled and pursues a pre-defined collective goal. We assume that, every day, the fleet operator decides how many CAVs will be routed via each alternative. Once this decision has been made the CAVs set off onto the prescribed routes and, during the process of driving, behave similarly to HDVs. In particular, we assume that CAVs do not utilize more efficient driving techniques such as platooning \cite{PlatoonReview, WangSurvey}. The only aspect differentienting CAVs from HDVs that we consider in this paper is collective route choice based on superior access to information about the system and prediction of human drivers' behaviour. Once the modified system has stabilized (in most cases) again, we compare the statistics of the system \emph{before} and \emph{after} the introduction of CAVs and reach our conclusions.
  
\noindent Let us note that similar frameworks under the name of guidance systems, Advanced Travel Information Systems (ATIS) or Stackelberg congestion games \cite{Harker, Rosenthal, VanVuren, YangATIS, Yang} have been considered in the literature. However, in contrast to them our goal is to demonstrate a range of outcomes with emphasis on the ordinary human driver as well as system-wide welfare when confronted with a centrally-guided fleet of CAVs rather than to show how the traffic system could be made more efficient or brought closer to system optimum, compare \cite{Kashmiri, ZhangNie}. 
Furthermore, we explicitly consider the decision process and gradual adaptation of human drivers as opposed to a typical Stackelberg game setting \cite{Yang} of a Cournot-Nash company with market power and individual rational price-takers.
We also treat human drivers as separate entities with different tastes who take time to adapt without aggregating them into a single User Equilibrium player which can instantaneously arrive at an optimal equilibrium assignment \cite{VanVuren}. Moreover, in contrast to the repeated game setting typical in reinforcement learning \cite{SB}, we assume that human agents only take myopic decisions to minimize the current perceived travel time without optimizing their long-term pay-offs. 
Finally, our point of view is distinct regarding the CAVs. Namely, the fleet of CAVs, even if it represents a robo-taxi company carrying people who switched from their own cars or caters for people who have subscribed to a collective route guidance system (like online routing services), is a separate entity with its own target which might diverge from the goal of city authorities or even be a reflection of hidden hostile motives. In this vein, we do not assume that the system necessarily stabilizes after introduction of CAVs. Indeed, for some fleet strategies, as we demonstrate, a stable state is an undesirable feature, and keeping the system away from it allows the fleet of CAVs to maximize its specific collective target, compare \cite{Akman} for more general multi-agent targets. 
Finally, the fleet has full information regarding the system travel times and can predict how many HDVs will choose every alternative before making their own routing decision, see also \cite{Akman, Psarou} for reinforcement learning-based city-scale scenarios.

\section{Results}
\label{Sec_Res}
\begin{figure*}[h]
\centering
\includegraphics[scale=0.45]{"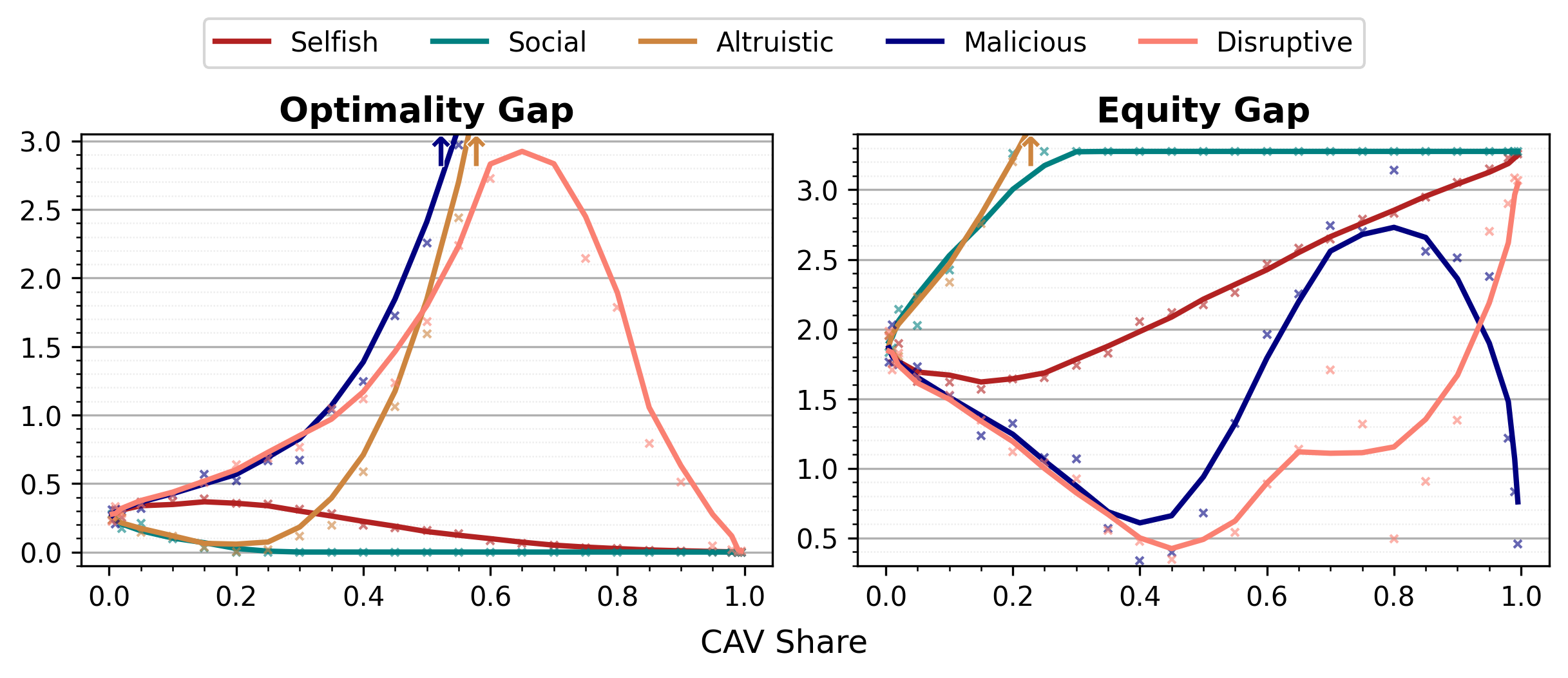"}
\caption{Optimality gap (distance from the system optimum, in which the mean travel time of all the drivers is lowest) 
and equity gap (standard deviation of travel times in the system) for different CAV shares and strategies. The optimality gap is $0$ for CAV shares large enough in the social strategy when the system reaches optimum, this however entails considerable equity gap. The selfish strategy is similar to the social one.
 The altruistic strategy results in large optimality and equity gaps for high CAV shares. Malicious and disruptive strategies exhibit varying optimality and equity gaps. }
\label{Fig_6}
\end{figure*}

\begin{figure*}[h]
\centering
\includegraphics[scale=0.45]{"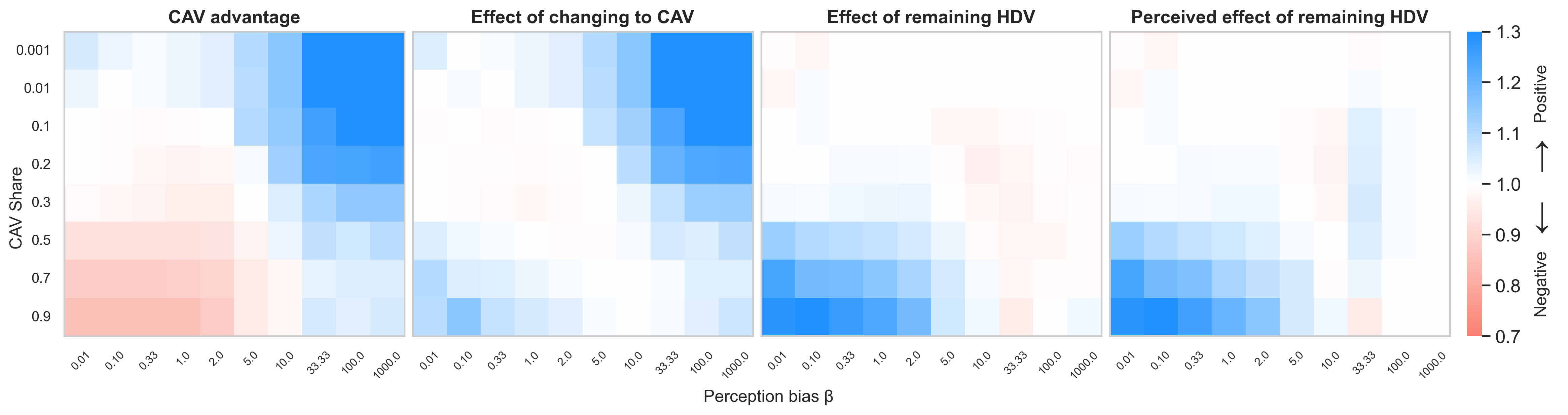"}
\caption{Positive and negative consequences of introduction of CAVs for the selfish CAV strategy and different fleet shares and spread of human preferences (perception bias), compare Fig. \ref{Fig_5}. CAV advantage (left) is particularly high for high human bias and low fleet shares. The situation is opposite for high fleet share and low spread. Effect of changing to CAV (middle-left) is virtually always beneficial. Effect of remaining HDV (middle-right) is beneficial primarily for low spread and high fleet shares. Otherwise it could be slightly negative (middle-right and right).} 
\label{Fig_7}
\end{figure*}

\begin{figure*}[t]
\centering
\includegraphics[scale=0.45]{"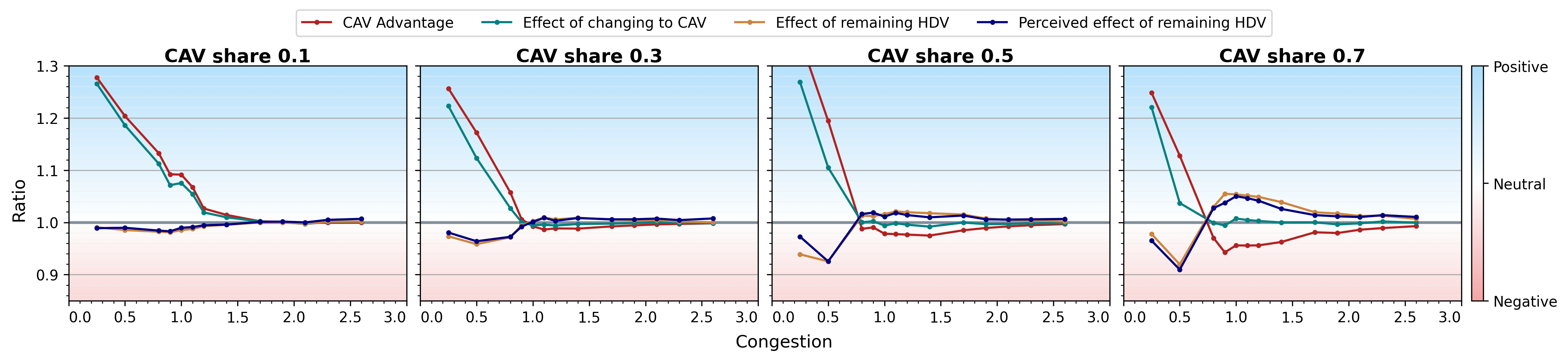"}
\caption{Outcomes of replacing a fraction of HDVs by CAVs for the selfish strategy and different congestion levels $C$ (where  $C * 1000$ is the total number of drivers in the system) based on CAV advantage, Effect of changing to CAV, Effect of remaining HDV and Perceived effect of remaining HDV, see Fig. \ref{Fig_5}. For small congestion levels the effect or remaining HDV is negative and this negativity deepens with increasing CAV share. Contrariwise, the effect of changing to CAV is positive. Very high congestion levels result in the agents being indifferent to whether they change into a CAV or remain HDV. 
For intermediate congestion, CAV advantage is negative and Effect of remaining HDV positive, the more so the larger the CAV share.} 
\label{Fig_8}
\end{figure*}

In the main experiment we study the long-term consequences of different proportions of HDVs becoming a centrally-coordinated fleet of CAVs in our two-route scenario. We compare the choices and travel times of HDVs and CAVs and summarize the results in Figs. \ref{Fig_3}, \ref{Fig_4}, \ref{Fig_5}, \ref{Fig_6}. In the second experiment, Fig. \ref{Fig_7}, we examine the dependence of the results on perception bias of human agents. Finally, in the third experiment, Fig. \ref{Fig_8}, we study how the results depend on congestion.  

\noindent{\bf Experimental setting}\\ 
In the experiments, run in a custom simulation software, we let the system composed of only  HDVs stabilize and, after $200$ days (on M-day) we replace a given share of HDVs by CAVs. We study the system purely experimentally in the stable regime of parameters summarized in Table \ref{Table_CoexParams}, see Appendix for experiments motivating this choice. For human drivers we assume the $\epsilon$-Gumbel model, see Methods. For CAVs, we consider five possible strategies (see Table \ref{Table_Goals} and Methods). After M-day, we run the simulation for another $100$ days, see Fig. \ref{Fig_2}, after which we record HDV and CAV travel times and flows (vehicle counts) on both routes and compare them to the respective values before M-day. 
We distinguish five phases:
\begin{itemize}
\item Days 1-100: Stabilization of HDV-only system composed of, by default, $1000$ drivers.
\item Days 101-200: Stable state in which we capture various statistics for HDVs.
\item Day 200 (M-day): a given share of HDVs is replaced by a centrally-coordinated fleet of CAVs.
\item Days 201 - 300: Stabilization of the system in the new reality.
\item Days 301 - 400: Stabilized (for most cases) state in which we compute the same statistics, this time for both HDVs and CAVs. We compare them to each other as well as to the statistics from days 101-200.
\end{itemize}

\noindent{\bf Are route choices of HDVs and CAVs similar?}

\noindent Figure \ref{Fig_4} shows that they are quite different. In the selfish scenario, for instance, the fraction of CAVs choosing A is $1$ for low CAV shares and it decreases to ca. $0.6$ for share $1.0$, which corresponds to system optimum. Fraction of HDVs choosing A, on the other hand, seems to increase for increasing CAV share. In the social case, we observe an exactly opposite tendency, with all CAVs selecting B for low CAV shares. In the altruistic case, virtually all CAVs select the route with longer travel time. In the malicious and disruptive cases, the differences between HDV and CAV choice patterns are also considerable. 

\noindent Overall, we conclude that the routing choices made by fleets of CAVs differ substantially from those made by HDVs accross various CAV strategies. This changes the system and affects HDVs' travel times.

\noindent{\bf Are HDVs better off after the introduction of CAVs? Are CAVs better off than HDVs?}

\noindent We consider the following statistics (see Methods):
\begin{itemize}
\item  mean travel time of HDVs averaged over days $101$-$200$, i.e. before introduction of CAVs ($\tau_b$),  
\item  mean travel time of HDVs averaged over days $301$-$400$ ($\tau$), 
\item mean perceived travel time of HDVs averaged over days $101$-$200$ ($u_b$).
\item  mean perceived travel time of HDVs averaged over days $301$-$400$ ($u$),
\item  mean travel time of CAVs averaged over days $301$-$400$ ($\rho$),
\end{itemize}
Studying the ratios $\tau$/$\rho$, $\tau_b$/$\rho$, $\tau_b$/$\tau$ and $u_b$/$u$  we discover that (Fig. \ref{Fig_5}):
\begin{itemize}
\item For modest CAV shares, CAVs are better off compared to HDVs before M-day (\emph{effect of changing to CAV} $>1$) in the selfish, malicious and disruptive scenarios, while HDVs are worse off (\emph{effect of remaining HDV} $<1$). Consequently, there seem to exist scenarios in which CAVs improve their total travel time at a cost to HDVs, see also Fig. \ref{Fig_8}. The effect is opposite in the social and altruistic strategies, where CAVs bear the cost of improving the driving conditions for HDVs or  for the entire system, compare Fig. \ref{Fig_6}. 

\item  Larger shares of CAVs render selfish and especially malicious and disruptive strategies costly to CAVs (\emph{CAV advantage} as well as  \emph{Effect of changing to CAV} drop below $1$). The high cost of altruistic strategy skyrockets while the social strategy becomes more and more cost-effective as CAV share increases. Larger CAV shares result also in oscillations in the system for malicious and disruptive strategies, confirmed by the bimodal distribution in Fig. \ref{Fig_3}, see also Suppl. Fig. 9 for more details. 

\item The influence of M-day on mean perceived travel times is very similar to the influence on actual travel times (the panels \emph{Effect of remaining HDV} and \emph{Perceived Effect of remaining HDV} are similar).
\end{itemize}

To summarize, CAVs and HDVs might be both better off and worse off compared to HDVs before introduction of CAVs and the outcome depends on the share and strategy of CAVs. 
In particular, for certain feasible combinations of parameters, the most disturbing scenario when CAVs gain and HDVs are disadvantaged may occur. 

\noindent{\bf Is the system closer to optimum?}\\
\noindent Fig. \ref{Fig_6} demonstrates that the social strategy reduces the optimality gap for the price of increased equity gap. The selfish strategy makes the system less efficient for low CAV shares and more efficient for large CAV shares. The altruistic strategy is very inefficient and inequitable and the same, to a lesser degree, applies to malicious and disruptive strategies. 

\noindent{\bf Does perception bias of HDVs influence the reaction of the system to introduction of CAVs?}\\
\noindent In this experiment we vary the spread of human preferences $\beta$. Small $\beta$ corresponds to unbiased human behaviour (choosing the predicted faster route) while large $\beta$ makes HDV choices more random because of large spread of subjective preferences. Considering only the selfish CAV strategy we conclude that, Fig. \ref{Fig_7}, 
\begin{itemize}
\item More biased (large $\beta$) HDV choices allow CAVs to decrease their travel time after M-day, especially for small shares of CAVs. The impact on HDVs' travel times tends to be slightly negative.
\item Less biased (small $\beta$) HDV choices in combination with the selfish CAV strategy result in improvement of driving conditions for both types of agents, with HDVs gaining the most. This is particularly visible for larger shares of CAVs. 
\end{itemize} 

\noindent{\bf Does congestion in the system influence the consequences of introduction of CAVs?}\\
In this experiment, we vary congestion levels, keeping human drivers' perception bias at the baseline level and letting CAVs apply the selfish strategy. We conclude that, Fig. \ref{Fig_8}:
\begin{itemize}
\item Modest congestion lets CAVs gain considerably in the selfish strategy at a cost to human drivers. The negative impact on HDV travel times increases as CAV share increases. Heavy traffic, on the other hand, makes the system more rigid, precluding any substantial gain in CAV travel time.  
\item In the intermediate congestion regime, CAVs are better off in terms of travel times than HDVs, with the effect more pronounced for higher CAV shares.  
\end{itemize}

\section{Discussion}
In this research we provide evidence for the existence of certain phenomena emerging from HDV-CAV interaction in the context of route choice which are of paramount significance for the performance of future urban systems. Our abstract models deliberately reduce the complexity of the problem in order to highlight these typical phenomena, which are likely to be even more pronounced in real urban mobility systems. To achieve this, we abstract reality at three main levels: network topology, traffic flow, human route-choice decision process.

\begin{itemize}
\item {\it Network.} In the complex networks of real megacities the number of available routes is huge and, consequently, the everyday action space for CAVs is much larger. It might include route alternatives not considered by humans in their choice process \cite{Gao}. We argue that if CAVs, like here, manage to identify effective strategies in two-route abstract networks, they are likely to identify them within more complex topologies. On the other hand, even very dense and complex urban road network topologies tend to have bottlenecks, 
where demand exceeds capacity. Consequently, the competitive strategy of route-choice is to exploit the capacity at isolated bottlenecks \cite{Geroliminis, Ji, Leclercq}, which might resemble the two route scenario considered in the paper. 

\item {\it Traffic flow.} The traffic flow, in reality non-deterministic, highly variable, controlled by static and adaptive traffic lights and with diverse microscopic phenomena such as platooning, accidents, slow-vehicles and driver errors \cite{Avila, Saberi} is only coarsely approximated with static BPR functions. We argue, however, that if machines manage to identify better routes in static analytical models with strictly increasing travel times (BPR), they will find even more opportunities in fluctuating, sensitive to outliers and non-continuous granular spatiotemporal patterns of real traffic flow \cite{Colak, Olmos}.

\item {\it Human decision process.} The long postulated Nash Equilibrium in traffic networks seems to be over-optimistic and  hardly observed empirically. 
Here, we applied machine actions in a system much more equilibrated than observed in the real world \cite{Zhu}, compare Appendix. 
We argue that, instead, real systems are ensembles of various  agents with different motives, utilities, capabilities and taste heterogeneities as in the seminal El Farol bar paper \cite{Arthur}. Consequently, the agents are much more diverse than the rigid utility maximisers considered here, which is likely to facilitate the task of CAVs. 
\end{itemize}

\noindent 
In all these aspects our models seem to be more restrictive for CAVs than real world and yet we were able to clearly reveal the disturbing phenomena. Hence, the results might be even sharper when, as it is in real cities, the network topology and traffic flow are complex, humans are even less optimal or homogenous and advanced machine learning is used to optimize CAV strategies. On the other hand, in the real world, human drivers might have better access to information facilitated by new technologies. Moreover, we modeled travel times by simple analytical BPR functions which are easily optimized by machines. In reality, CAVs will not have such precise information about the system and their optimization is likely to be based on reinforcement learning \cite{Akman, Psarou} and high performance computing.

\noindent The advantages of CAVs visible in our experiments can be summarized as follows. 
\begin{itemize}
\item Advantage by collective decision taking, e.g. strategies that improve the average travel time of the fleet or of the system, which are hardly possible if every agent, like humans, is independent. 
\item Advantage by better access to information and information sharing, e.g. perfect understanding of the characteristics of the traffic system. 
\item Advantage by advanced processing and optimization capabilities, e.g. human behaviour modeling, human choice prediction. 
\item Advantage by lack of perception error, i.e. 
decisions based on \emph{actual} as opposed to \emph{perceived} travel times.
\item Advantage by instantanous adaptation, which allows the machines to keep the system out of equilibrium and exploit slower human drivers' adaptation, as is the case for malicious and disruptive strategies, see also Appendix.
\end{itemize}
These sources of advantage enable more efficient CAV routing decisions. In the default selfish case the CAVs outperform human drivers by selecting on average faster routes 
 for small CAV market shares, Figs. \ref{Fig_3}, \ref{Fig_5}. For large market shares, the impact is more complex. Namely, the CAVs obtain better travel times than travel times of HDVs they replaced, however the driving conditions for HDVs improve even more, Fig. \ref{Fig_2}. This is due to the fact that they bring the system closer to optimum which involves different travel times on routes. The tipping poing is around $25\%$, Fig. \ref{Fig_5}, for the default moderate congestion levels and spread of human preferences.  System-wise, collective strategies of CAVs, even if they are selfish may reduce the mean travel time (see optimality gap, Fig. \ref{Fig_6}), reducing e.g. $CO_2$ emissions and noise \cite{Choudhary}. Other strategies, notably malicious and altruistic for large CAV shares, may increase the optimality gap, reducing the liveability of cities and sustainability of urban driving.

\noindent To summarize, CAV fleets will transform urban traffic systems. One of the aspects in which this will manifest itself will be route choice. The impact on the human drivers and urban welfare will depend on the strategies CAVs are allowed to adopt.
For instance, for the outright malicious CAV fleet strategy, the driving conditions will deteriorate for everyone. At the other end of the spectrum, the altruistic strategy might bring huge benefits to the HDVs which remain in the system.  

\noindent Non-standard strategies aside, however, our results indicate that even in the most straightforward scenarios with modest shares of CAVs minimizing their collective travel time the remaining human drivers might become disadvantaged as a side-effect. Do we want this?

\section{Methods}
\begin{table*}[h]
    \centering
\caption{CAV fleet optimization targets used in our experiments}
    \vspace{1ex}
    \begin{tabular}{|l|c|c|c|}
    \hline
      $\lambda^{CAV}$ & $\lambda^{HDV}$ & Interpretation & Optimization target\\
      \hline
      $1$ & $0$ & Selfish & minimize only CAV travel time\\
      $0$ & $1$ & Altruistic & minimize only HDV travel time\\
      $0$ & $-1$ & Malicious & maximize HDV travel time\\
      $1$ & $-9$ & Disruptive & maximize travel time for HDV and minimize for CAV \\
      $1$ & $1$ & Social & minimize total travel time (system optimum)\\
\hline
\end{tabular}
 \label{Table_Goals}
\end{table*}

We run agent-based simulations, where human drivers are modelled as independent heterogenous rational utility maximizers who learn from experience to maximize expected utility. They share the network, where congestion is modelled with a static BPR function, with CAVs, whose behaviour in the traffic is the same as HDVs', except for routing. The focus is on representing collective route choices of CAVs while HDVs' choice is based on adaptation of standard models well-established in the literature, see Introdution. 
 
\noindent{\bf Traffic Networks}

\noindent We abstract the traffic network to two independent non-overlapping routes, $A$ and $B$,  connecting one pair Origin-Destination. The travel times are functions of flow, given by the static BPR-type function \cite{BPR},  
\begin{equation}
\label{eq_BPR}
t_r(q_r) =   t_r^0 \left( 1 + \left(\frac {q_r} {Q_r}\right)^{b}\right),
\end{equation}
where $t_r^0$, for $r \in \{A, B\}$, is the free-flow travel time, which a traveller would experience travelling on an empty road. $Q_r$ is the capacity of the road section and $b>1$. Finally, $q_r$ is the number of vehicles choosing route r within a given interval of time such as $1$ hour. 
  
\noindent In our setting, we assume $b = 2$, the default total number of drivers, $q = q_A + q_B = 1000$ and the alternative routes' capacities and free-flow travel times are given by $t_A^0 = 5 \mbox{ min}$, $t_B^0 = 15 \mbox{ min}$, $Q_A = 500$, $Q_B = 800$, which corresponds to a shorter route (A) with low capacity and a longer route (B) with higher capacity. Note that when the number of cars on a given alternative exceeds its capacity the travel time rises steeply as in reality \cite{Geroliminis}.

\noindent{\bf Human Learning}\\
Let $T_A(i), T_B(i)$ denote the predictions (estimates) by human agent $i$ of travel times on routes $A$ and $B$, respectively. Suppose that, on a given day, agent $i$ travelled along route $r(i)$ and experienced travel time $t_{r(i)}$. Then the predicted travel times are adjusted by
\begin{equation}
\label{eq_ExpOnly}
T_r(i) \leftarrow \begin{cases} 
(1-\alpha)T_r(i) + \alpha t_{r(i)} &\mbox{ if } r = r(i), \\
T_r(i) &\mbox{ otherwise.}
\end{cases}  
\end{equation}
Above, $\alpha \in [0,1]$ is the learning rate which specifies the relative weight of the most recent experience. For a typical value of $\alpha = 0.2$, the new estimate of travel time is made up of $20\%$ most recent experience and $80\%$ previous estimate. Crucially, the estimate of travel time along the unused alternative remains unaltered.

\noindent{\bf Human Choice}\\
The human agents make choices according to the $\epsilon$-Gumbel model, which we introduce for our setting. In this model, following equation \eqref{Eq_Ur}, we assume that
\begin{equation}
U_r(i) = - T_r(i) + \varepsilon_r(i)
\label{Eq_UtH}
\end{equation}
is the perceived utility of alternative $r$ to agent $i$.   Predicted travel times $T_r(i)$ are updated daily by formula \eqref{eq_ExpOnly} while $\varepsilon_r(i)$ is a fixed real number sampled once, independently for every agent $i$ and alternative $r$, from distribution $Gumbel(\mu, \beta)$ with scale $\beta$ and location $\mu = - (\beta * \gamma_E)$. Here, $\gamma_E$ is the Euler-Mascheroni constant and the cumulative distribution function of $Gumbel(\mu, \beta)$ is given by $\exp(-\exp ((x-\mu)\beta^{-1}))$. The mean of $Gumbel( - (\beta * \gamma_E), \beta)$ is $0$ and variance equals $\frac{\pi^2\beta^2}{6}$. 

\noindent In the $\epsilon$-Gumbel model every agent has fixed, pre-specified tastes related to the alternatives, which are independent of the tastes of other agents. The larger $\beta$, which we call spread or bias, the more subjective, on average, the decisions of human agents. These decisions are based on maximizing utility \eqref{Eq_UtH} up to small exploration $\epsilon$ via formula:

\begin{equation}
r(i) = \begin{cases} 
\arg \min_{r \in \{A,B\}} U_r(i)  & 
\mbox{ with probability } 1-\epsilon,\\
\mbox{uniformly random}  & \mbox{ with probability } \epsilon.
\end{cases}
\label{eq_exploration}
\end{equation}
Once on a given day every agent, including both HDVs and CAVs (if there are any) has made its choice, we determine $q_A$ and $q_B$ as the total number of agents choosing $A$ and $B$, respectively, and use formula \eqref{eq_BPR} to compute travel times $t_A$ and $t_B$. We feed these values into equation \eqref{eq_ExpOnly} to update HDV estimates on the following day and close the HDV experience-learning-choice loop by incrementing the day number, see Fig. \ref{Fig_2}.  
\begin{table*}[h]
\centering
\caption{Parameters used in the experiments on coexistence of HDVs and CAVs}
\vspace{1ex}
\begin{tabular}{ |l|c|c| }
  \hline
	Parameter & Default Value & Alternative Values\\
  \hline
	HDV Model Type & $\epsilon$-Gumbel & - \\
	Human perception spread ($\beta$) & $5.0$ &  $0.01$ -- $1000.0$\\
	Initial HDV Knowledge & Free Flow & - \\
	Initial HDV Choice & Random & -\\
	HDV learning rate ($\alpha$) & $0.2$ & - \\
	HDV exploration rate ($\epsilon$) & $0.1$ & -\\
	HDV Learning From Experience Only & True & -\\
	CAV optimization strategy & Selfish & Malicious, Disruptive, Altruistic, Social\\
	CAV share & $0.0$ & $0.0 - 1.0$\\
	Congestion & $1.0$ & $0.25$ -- $2.6$	\\
  \hline
\end{tabular}
\label{Table_CoexParams}
\end{table*}

\noindent{\bf CAV Optimization}\\
\noindent We assume that CAVs optimize their pre-defined target taking advantage of their superior knowledge about the system and human agents' decisions. Namely, on a given day, before deciding how many CAVs to route via $A$ and $B$, the fleet operator predicts perfectly the numbers of HDVs, $q^{HDV}_A$, $q^{HDV}_B$ which intend to travel on $A$ and $B$, respectively. Then, it selects the number $q^{CAV}_A$ of agents it routes via $A$ such that $0 \le q^{CAV}_A \le q^{CAV}$, where $q^{CAV}$ is the total number of centrally controlled machine agents, and $q^{CAV}_A$ minimizes the target function
\begin{equation*}
\Phi_{\lambda}({q^{CAV}_A}) := \lambda^{CAV} t^{CAV} + \lambda^{HDV}t^{HDV}, 
\end{equation*}
where
\begin{eqnarray*}
t^{CAV} &=& {q^{CAV}_A} t_A(q_A) + q^{CAV}_B t_B(q_B),\\ 
t^{HDV} &=& q^{HDV}_A t_A(q_A) + q^{HDV}_B t_B(q_B), \\
q^{CAV}_B &=& q^{CAV} - {q^{CAV}_A},\\
q_A &=& q^{HDV}_A + {q^{CAV}_A},\\
q_B &=& q^{HDV}_B + q^{CAV}_B,
\end{eqnarray*}
and $t_A(q_A)$, $t_B(q_B)$ are given by \eqref{eq_BPR}. Coefficients $\lambda^{CAV}$, $\lambda^{HDV}$ depend on the strategy adopted by CAVs, see Table \ref{Table_Goals} and Fig. \ref{Fig_2}.

\noindent{\bf Parameters}\\
Table \ref{Table_CoexParams} presents the parameters used to study the coexistence of HDVs and CAVs. The default human choice model is the $\epsilon$-Gumbel model described above, see Appendix for alternatives. Parameter $\beta$, by default equal to $5.0$, accounts for reasonable spread between the alternatives. Nevertheless, we vary it in the range $0.01$ - $1000.0$, Fig. \ref{Fig_7}, which allows us to test the systems for very unbiased HDVs, whose utility is very close to predicted travel times, as well as systems where human tastes and, consequently, decisions look random to an external observer. Initial HDV Knowledge and Initial HDV Choice account for the initial conditions in the simulation on day $1$. We assume that for every human agent $i$, $T_r(i) = t_r^0$ for $r \in \{A, B\}$ on day $1$ (Free Flow Initial Knowledge) and the first choice $r(i)$ (Initial HDV Choice) is Random. We note that the first choice does not significantly influence the outcomes of the simulations, see Appendix. HDV learning and exploration rates equal $0.2$ and $0.1$, respectively, and we assume that HDVs learn from experience only. 
Finally, the total number of vehicles is equal to $1000 * C$, where the default congestion, $C$, is set to $1.0$ resulting in $1000$ agents. This level amounts to $77 \%$ of the total capacity of the system, equal to $500 + 800 = 1300$ and is moderate. However, we consider the traffic congestion from very light ($0.25$) up to gridlocked ($2.6$), Fig. \ref{Fig_8}.

\noindent{\bf Statistics used in experiments}\\
Here we assume that before M-day there are $1000$ HDVs in the system (for congestion $C$ different from baseline the formulas are adjusted accordingly). 
As the characteristics of the HDVs are assigned randomly, we assume that the HDVs that are replaced by CAVs are the HDVs with the last $1000*shareCAV$ indices. Therefore, after M-day $q^{CAV} = 1000*shareCAVs$. Let us also denote $q^{HDV}_* := 1000 - q^{CAV}$. 

The statistics we report in our experiments are the following. After a given day of simulation we compute:
\begin{itemize}
\item Mean travel time of HDVs: $$\frac {1}{q^{HDV}}\sum_{i=1}^{q^{HDV}} t_{r(i)},$$ where $t_{r(i)}$ is the travel time experienced by agent $i$ on a given day. Importantly, the number of HDVs, $q^{HDV}$ is not constant and is equal to $1000$ until M-day and $q_*^{HDV}\le 1000$ after $M$-day. 
\item Mean perceived travel time of HDVs: $$\frac {1}{q_*^{HDV}}\sum_{i=1}^{q_*^{HDV}} (t_{r(i)} + \varepsilon_{r(i)}(i)),$$ 
Note that in contrast to mean travel time of HDVs, the mean perceived travel time before M-day is computed only for the vehicles that remain HDVs after M-day. 
\item Mean travel time of CAVs:  $$\frac {1}{q^{CAV}}\left(q^{CAV}_A t_A(q_A) + q^{CAV}_B t_B(q_B)\right).$$
\item Fractions of HDVs and CAVs on A: $q_A^{HDV}/q^{HDV}$ and $q_A^{CAV}/q^{CAV}$, respectively.
\end{itemize}
These one-day statistics are then averaged over days $101$-$200$ or $301$-$400$ to $\tau_b, \tau, u_b, u, \rho$ and average fractions, see Results. The system-wide statistics are:
\begin{itemize}
\item Optimality gap: ${S} - {S_O}$ averaged over days $301-400$, where $$S = \frac {q_A t_A(q_A) + q_B t_B(q_B)}{q_A + q_B}$$ is the mean travel time of all vehicles on a given day and $S_O = \min_{q_A} S$ is the least possible mean travel time (experienced in System Optimum).
\item Equity gap: $\sigma$ averaged over days $301$ - $400$, where $$\sigma = \sqrt{ \frac {q_A(t_A(q_A) - S)^2 + q_B(t_B(q_B) - S)^2}{q_A + q_B}}$$ is the standard deviation of travel times of all the drivers on a given fixed day. 
\end{itemize}
\noindent{\bf Reproducibility}\\
To verify reproducibility of the core findings, we reran the main experiment $10$ times for selected CAV shares (those used in Fig. \ref{Fig_3}). We obtained statistical significance of the results presented in Fig. \ref{Fig_2} (with $p<0.001$) using the paired two-tailed t-test with nine degrees of freedom, see Appendix. 

\section{Code and data availability}
The experiments were performed using custom light-weight simulation software, BottleCOEX, available online as a github repository at \url{https://github.com/COeXISTENCE-PROJECT/BottleCOEX} along with the supplementary information (Appendix) and the data used for the experiments described in this paper.   
\\
\section{Author contributions}
GJ - conceptualization, methodology, software, validation, data analysis, writing - original draft, writing - review and editing; AOA - visualization, software, data analysis, writing - review and editing; AP - visualization, software, data analysis, writing - review and editing; ZGV - visualization, data analysis, writing - review and editing; RK - conceptualization, methodology, visualization, data analysis, writing - review and editing, project administration, funding acquisition.

\section{Competing interests}
The authors declare no competing interest. 

\section{Acknowledgement}
This research was supported by the ERC Starting Grant number 101075838: COeXISTENCE.

\end{document}